\documentclass[journal,10pt]{IEEEtran}
\IEEEoverridecommandlockouts
\usepackage{amsmath,amssymb,amsfonts}
\usepackage{algorithmic}
\usepackage{graphicx}
\usepackage{textcomp}
\usepackage{xcolor}
\usepackage[ruled, linesnumbered]{algorithm2e}
\usepackage{textcomp}
\usepackage{booktabs}
\usepackage{multirow}
\usepackage[numbers,square,comma,compress,sort]{natbib}
\def\BibTeX{{\rm B\kern-.05em{\sc i\kern-.025em b}\kern-.08em
    T\kern-.1667em\lower.7ex\hbox{E}\kern-.125emX}}

\newcommand{\rev}[1]{\textcolor{black}{#1}}
\newcommand{\esl}[1]{\textcolor{black}{#1}}
\newcommand{\camready}[1]{\textcolor{black}{#1}}

\usepackage{color, colortbl}
\definecolor{LightCyan}{rgb}{0.88,0.9,1}
\definecolor{Burgundy2}{RGB}{158,5,8}
\newcommand{\birH}{{\textit{\textcolor{black}{H}}}}

\usepackage{changepage}

\begin{document}

\title{\esl{Analytical Performance Modeling of NoCs under \\ Priority Arbitration and Bursty Traffic\vspace{-2mm}}}

\author{Sumit K. Mandal$^1$, Raid Ayoub$^2$, Michael Kishinevsky$^2$, Mohammad M. Islam$^2$, Umit Y. Ogras$^1$ \\
$^1$School of ECEE, Arizona State University; 
$^2$Intel Corporation, Hillsboro, OR \vspace{-10mm}
\thanks{This work was supported partially by Strategic CAD Labs, Intel Corporation, USA.

S. K. Mandal {and} U. Y. Ogras, School of
Electrical,
Computer and Energy Engineering, Arizona State University, Tempe, AZ,
85287; emails: \{skmandal, umit\}@asu.edu;

Raid Ayoub, Michael Kishinevsky {and} Mohammad M. Islam, Intel Corporation, 2111 NE 25th Ave.,
Hillsboro, OR 97124; emails: \{raid.ayoub, michael.kishinevsky, mohammad.majharul.islam@intel.com\}@intel.com
}}

\maketitle


\begin{abstract}

\rev{Networks-on-Chip (NoCs) used in commercial many-core processors typically incorporate priority arbitration. Moreover, they experience bursty traffic due to application workloads.}
However, most state-of-the-art NoC analytical performance analysis techniques assume fair arbitration and simple traffic models. 
To address these limitations, we propose an analytical modeling technique for priority-aware NoCs under bursty traffic.
Experimental evaluations with synthetic and bursty 
traffic 
show that the proposed approach has less than 10\% modeling error with respect to cycle-accurate NoC simulator.


\end{abstract}

\vspace{-4mm}
\section{Introduction}
\vspace{-1mm}

Industrial many-core processors incorporate priority arbitration for the routers in NoC~\cite{jeffers2016intel}.
Moreover, these designs execute bursty traffic since real applications exhibit burstiness~\cite{bogdan2010workload}.
Accurate NoC performance models are required to perform design space exploration and 
accelerate full-system simulations~\cite{kiasari2013analytical, qian2015support}.
Most existing analysis techniques assume fair arbitration in routers, \rev{which does not hold for NoCs with priority arbitration used in
manycore processors, such as high-end servers~\cite{tam2018skylake} and high performance computing (HPC)~\cite{jeffers2016intel}}.
\rev{A recent technique targets priority-aware NoCs~\cite{mandal2019analytical}, 
but it assumes that the input traffic follows geometric distribution.} 
While this assumption simplifies analytical models, \textit{it fails to capture the bursty behavior of real applications}~\cite{bogdan2010workload}. 
Indeed, our evaluations show that the geometric distribution assumption leads up to 60\% error in latency estimation unless the bursty nature of applications is explicitly modeled. 
Therefore, there is a strong need for NoC performance analysis techniques that consider both priority arbitration and bursty traffic.



This work proposes a novel performance modeling technique for priority-aware NoCs that takes bursty traffic into account. 
It first models the input traffic as a generalized geometric (GGeo) discrete-time distribution that includes a parameter for burstiness. 
We achieve high scalability by employing the principle of maximum entropy (ME) to transform the given queuing network into a near equivalent set of individual queue nodes of multiple-classes with revised characteristics (e.g., modifying service process). Furthermore, our solution involves transformations to handle priority arbitration of the routers across a network of queues. Finally, we construct analytical models of the transformed queue nodes to obtain end-to-end latency.
\rev{The proposed performance analysis technique is evaluated with SYSmark\textsuperscript{\textregistered} 2014 SE~\cite{SYSmark2014}, applications from SPEC CPU\textsuperscript{\textregistered} 2006~\cite{henning2006spec} and SPEC CPU\textsuperscript{\textregistered} 2017~\cite{bucek2018spec} benchmark suites, as well as synthetic traffic. The proposed technique has less than 10\% modeling error with respect to an industrial cycle-accurate NoC simulator.}
The major contributions of this work are as follows:
\begin{itemize}
    \item Accurate and scalable high-level performance modeling of priority-based NoCs considering burstiness,
    \item Dynamic approximation of realistic bursty traffic via GGeo distribution,  
    \item Thorough evaluations on industrial priority-based NoCs with  synthetic traffic and real applications.
\end{itemize}
\vspace{-4mm}
\section{Related Work}
\vspace{-1mm}


NoC analytical performance analysis techniques primarily target fast design space exploration and accelerating full-system simulations.
Most of the existing techniques consider NoC routers with fair arbitration~\cite{ogras2013modeling,qian2015support}, 
but this assumption does not hold for NoCs that employ priority arbitration~\rev{\cite{jeffers2016intel,tam2018skylake}}.
%
Several performance analysis techniques target priority-aware NoCs~\cite{kiasari2013analytical, mandal2019analytical}.
The technique presented in~\cite{kiasari2013analytical} assumes that each class of traffic in the NoC occupies different queues.
This assumption is not practical since most of the industrial NoCs share queues between multiple traffic classes.
Analytical model for industrial NoCs, which estimates average end-to-end latency is proposed  in~\cite{mandal2019analytical}.
However, these models assume that the input traffic follows geometric distribution, 
which is not applicable for workloads with bursty traffic. 

Analytical modeling of priority-based queuing networks has also been studied outside of the realm of the on-chip interconnect~\cite{bolch2006queueing, walraevens2004discrete}. 
Analytical models constructed in~\cite{bolch2006queueing} considers a queuing network in the continuous-time domain.
This assumption is not valid for NoCs, as events happen in discrete clock cycles.
In~\cite{walraevens2004discrete}, performance analysis models are constructed in the discrete-time domain.
Since the number of random variables required in this technique is equal to the number of classes (exponential on the number of routers) present in the NoC, this approach does not scale. 
In contrast, 
the analytical models presented in this paper use the discrete-time domain and scale to thousands of traffic classes.

\vspace{-2mm}
\section{Background and Overview}




The goal of this work is to construct accurate performance models for industrial NoCs under \textit{priority-arbitration} and \textit{bursty traffic}. 
\rev{We mainly target manycore processors used in servers, HPC, and high-end client CPUs~\cite{jeffers2016intel,tam2018skylake}.}
The proposed technique takes burstiness and injection rate of the traffic as input and then provides end-to-end latency of each traffic class. 



\begin{figure}[t]
	\centering
	\vspace{-9mm}
	\includegraphics[width=0.7\columnwidth]{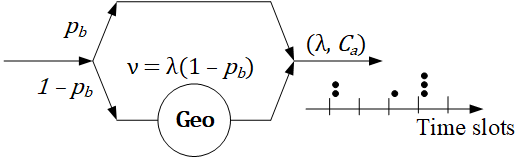}
	\vspace{-4mm}
	\centering
	\caption{GGeo traffic model} 
	\label{fig:traffic_model_GGeo}
    \vspace{-3mm}
\end{figure}

\vspace{1mm}
\noindent\textbf{Input traffic model assumptions: }Applications usually produce bursty NoC traffic with varying inter-arrival times~\cite{bogdan2010workload, qian2015support}.
We approximate the input traffic using the GGeo discrete-time distribution model, which takes both burstiness and discrete-time feature of NoCs into account~\cite{kouvatsos1994entropy,qian2015support}. 
GGeo model includes Geometric and null (no delay) branches, as shown in Figure~\ref{fig:traffic_model_GGeo}. 
Selection between branches conforms to the Bernoulli trial, where the null (upper) and Geo (lower) branches are selected with probability $p_b$ and $1-p_b$, respectively.
The Geo branch leads to geometrically distributed inter-arrival time, while the null branch issues additional flit in the current time slot leading to a burst. 
Both the number of flits in a time slot 
and the inter-arrival rate
depend on $p_b$~\cite{kouvatsos1994entropy}.  
Hence, we use $p_b$ as a parameter of burstiness. 
GGeo distribution has two important properties~\cite{kouvatsos1994entropy}. \textit{First, } it is pseudo-memoryless, i.e. the remaining inter-arrival time is geometrically distributed. \textit{Second,} it can be described by its first two moments ($\lambda$, $C_a$), where $C_a^2 = 2/(1 - p_b) -\lambda - 1$.
We exploit these properties to construct analytical models.

\vspace{-4mm}
\section{\esl{Systematic Generation of Analytical Models}} \label{sec:model_gen}
\vspace{-1mm}
In industrial NoCs, flits already in the network have higher priority than new injections to achieve predictable latency~\cite{jeffers2016intel}.
This leads to nontrivial timing dependencies between the multi-class flits in the network. 
Hence, we propose a systematic approach for accurate and scalable performance analysis.
\rev{We note that the proposed technique can be extended to NoCs with fair arbitration if we assume that all classes have the same priority. However, we do not focus on non-priority NoCs since this domain has been studied in the past~\cite{ogras2013modeling}.
}



\vspace{-4mm}
\subsection{\esl{Maximum entropy for queuing networks}} \label{sec:max_entropy}
\vspace{-1mm}

We apply the principle of ME to queuing systems to find the probability distribution of desired metrics (e.g., queue occupancy)~\cite{kouvatsos1994entropy}. According to this principle, the selected distribution should be \textit{the least biased} among all feasible distributions satisfying the prior information in the form of mean values. 
The optimal distribution is found by maximizing the corresponding entropy function: 
we formulate a nonlinear programming problem and 
solve it analytically via the Lagrange method of undetermined multipliers as discussed next.

\begin{table}[b]
\vspace{-3mm}
\caption{Summary of the notations used in this paper}
\centering
\vspace{-3mm}
\label{tab:symbols_used}
\scriptsize{
\begin{tabular} {@{}ll@{}}
\hline
$\lambda, \lambda_m$       & Mean arrival rate of total traffic and class m                                         \\ \hline
$p_b$       & Probability of burstiness                                                                                                    \\ \hline
$T_m$, $\widehat{T}_m$             & Original and modified mean service time of class m flits                                                                                                    \\ \hline
$R, R_{mk}$          & \begin{tabular}[c]{@{}l@{}} Total residual time and residual time of class m while class k is served\end{tabular} \\ \hline                                                                                                                     
$\rho_m$          & Mean server utilization of class m flits (=$\lambda_m T_m$)                                                                             \\ \hline
$C_a$, $C_{a_m}$  & \begin{tabular}[c]{@{}l@{}}Coeff. of variation of interarrival time of total traffic and class m flits\end{tabular}   \\ \hline
$C_{s_m}, \widehat{C}_{s_m}$    & \begin{tabular}[c]{@{}l@{}} Coeff. of variation of original and modified service time of class m flits\end{tabular} \\ \hline
$C_d$, $C_{d_m}$  & \begin{tabular}[c]{@{}l@{}}Coeff. of variation of interdeparture time of total traffic and class m flits \end{tabular} \\ \hline
$W_m$             & Mean waiting time of class m flits                                                                                                      \\ \hline
$\overline{n}_m, n_m$  & Mean and current occupancy of class m flits in a queue-node                                                                                                         \\ \hline
$\beta_m $        & Mean number of bursty arrivals of class m\\ \hline    
$\overline{n}_{mk}$ & Mean queue-node occupancy of class $m$ with serving class $k$\\ \hline
$\mathbf{n}$      & State vector, $\mathbf{n} = (n_1, n_2, ...,n_M)$ of priority queue-nodes\\ \hline
$p(\mathbf{n})$   & Probability that a queue-node is in state $\mathbf{n}$\\ \hline
$p_m(0)$          & Marginal probability of zero flits of class m in a queue-node.\\ \hline 
$\alpha_m(\mathbf{n})$ & $\alpha_m(\mathbf{n}) = 1$ if class $m$ in service and 0 otherwise \\ \hline
$M$ & Number of classes that share same server\\ \hline

\end{tabular}
}
\vspace{-6mm}
\end{table}
\vspace{-4mm}
\subsection{\esl{Decomposition of basic priority queuing}} \label{sec:basic_prior}
\vspace{-1mm}
\label{basic_priority_decomp}

\begin{figure}[t]
	\centering
	\vspace{-10mm}
	\includegraphics[width=0.85\columnwidth]{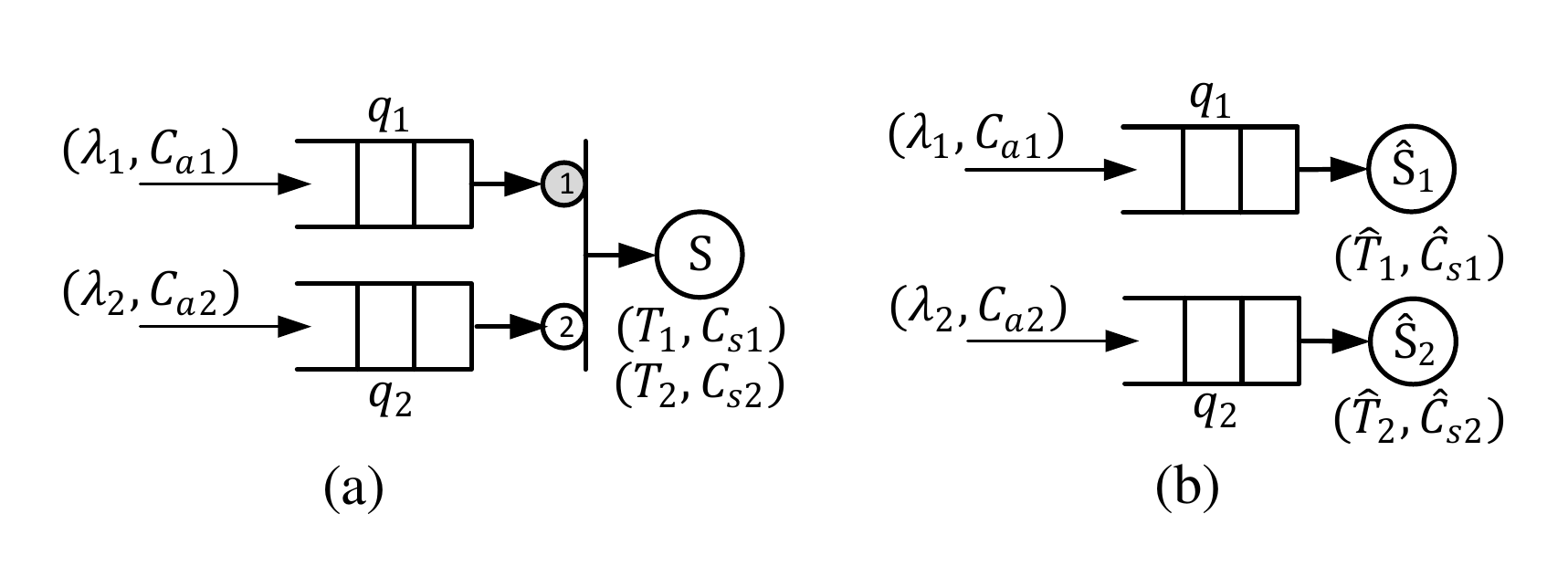}
	\vspace{-7mm}
	\centering
	\caption{Decomposition of a basic priority queuing} 
	\label{fig:two_class_prioirty_decomp}
	\vspace{-4mm}
\end{figure}

In a non-preemptive priority queuing system, the router does not preempt a higher priority flit while processing a lower priority flit. 
An example system with two queues and a shared server is shown in Figure \ref{fig:two_class_prioirty_decomp}(a). 
There are two flows arriving at a priority-based arbiter and a shared server. 
The shaded circle corresponds to high priority input (class 1) to the arbiter. We denote this structure as {\em basic priority queuing}.
Our goal is to decompose this system into individual queue-nodes with modified servers, as shown in Figure \ref{fig:two_class_prioirty_decomp}(b). 
The combination of a queue and its corresponding server 
is referred to as a {\em queue-node}. 
The effective expected service time of class 2 flits, $\hat{T}_2$, 
is larger than the original mean service time $T_2$, 
since class 2 flits wait for the higher priority (class 1) flits 
in the original system. 
We calculate the effective service time in the transformed network using Little's Law as:
\vspace{-0.5mm}
\begin{equation} \label{eq:t_m}
    \widehat{T}_m = \frac{1 - p_m(0)}{\lambda_m} 
\end{equation} 
where $p_m(0)$ is the marginal probability of having no flits of class $m$ in the queue-node,
as listed in Table~\ref{tab:symbols_used}. 

\noindent\textbf{\rev{Computing $p_m(0)$ using ME:}} We find $p_m(0)$ using the ME principle by maximizing the entropy function $H(p(\mathbf{n}))$ given in (\ref{eq:entropy}) subject to the constraints
listed in (\ref{eq:max_entropy}):
%
%
\begin{align} \label{eq:entropy}
& \underset{p}{\textrm{maximize}} && ~~H(p(\mathbf{n})) = -\sum_{\mathbf{n}} p(\mathbf{n})\log(p(\mathbf{n}) \\  \nonumber%
& \textrm{subject to} && ~~\sum_\mathbf{n=0}^{\mathbf{\infty}} p(\mathbf{n}) = 1, \\ \label{eq:max_entropy} 
 & & & ~\sum_{\substack{\mathbf{n} = \mathbf{0} \\ \text{except} \\ n_m = 1}}^{\mathbf{\infty}} \alpha_m(\mathbf{n})p(\mathbf{n}) = \rho_m, ~m = 1, \ldots, M\\ \nonumber
& & &  \sum_{\substack{\mathbf{n} = \mathbf{0}  \\ \text{except} \\ n_m = n_k = 1}}^{\mathbf{\infty}} n_m \alpha_k(\mathbf{n}) p(\mathbf{n}) = \bar{n}_{mk},
m,k=1,.., M
\end{align}
The notation $\mathbf{\infty}$ means a state vector $\mathbf{n}$ with all elements set to $\infty$, and ($\mathbf{n} = \mathbf{0} \text{ except } n_m = 1$) refers to a vector $\mathbf{n}$ with the $m^\textrm{th}$ element set to 1 and other elements set to 0. 
The constraints in (\ref{eq:max_entropy}) comprise three types:
normalization, mean server utilization 
and mean occupancy.
We introduced an extended set of mean occupancy constraints compared to~\cite{kouvatsos1994entropy} to provide further information about the underlying system. When a flit of a certain class arrives at the system, it may find the server busy with its own class or other classes since the server is a shared resource, as shown in Figure~\ref{fig:two_class_prioirty_decomp}(a). Therefore, the mean occupancy of each class can be partitioned according to the contribution of each class occupying the server. We exploit this inherent partitioning to generate $M$ additional occupancy constraints. 
The occupancy related constraints depend on three components, $\beta_m$, $R_{mk}$ and $W_m$ (defined in Table~\ref{tab:symbols_used}) 
derived in~\cite{kouvatsos1994entropy, mandal2019analytical}.
%

We solve the nonlinear programming problem in (\ref{eq:entropy}, \ref{eq:max_entropy}) to find $p(\mathbf{n})$ which we use to determine the probability of having zero flits of class $m$, $p_m(0)$. 
The convergence of this solution is guaranteed when the queuing system is in a stable region. We derived the general expression for $M$ queues in a priority structure with a single class per queue as:
\begin{equation}
    p_m(0) = 1 - \rho_m - \sum_{k = 1, k\neq m}^{M}\rho_k\frac{\overline{n}_{mk}}{\rho_k + \overline{n}_{mk}}
    \label{equ:p_m(0)}
\end{equation}
%
Plugging the expression of $p_m(0)$ from (\ref{equ:p_m(0)}) into (\ref{eq:t_m}), we obtain the first moment of the service process.

\noindent\textbf{\rev{Computing second moment of the service time:}} Since we also need the second moment to characterize the GGeo traffic, we calculate the modified squared coefficient of variation of the service time for class $m$ ($\widehat{C}_{s_m}^2$).
We utilize the queuing occupancy formulation of GGeo/G/1~\cite{kouvatsos1994entropy} and the modified server utilization $\widehat{\rho}_m = \lambda_m\widehat{T}_m$ to obtain the following expression for $\widehat{C}_{s_m}^2$:
%
\begin{equation} \label{eq:coeff_serv}
    \widehat{C}_{s_m}^2 = \frac{(1 - \widehat{\rho}_m)(2\overline{n}_{m} - \widehat{\rho}_m) - \widehat{\rho}_m C_{a_m}^2}{{\widehat{\rho}_m}^2}  
\end{equation}

%
%

\vspace{-4mm}
\subsection{Decomposition of priority queuing with partial contention} \vspace{-1mm}
\vspace{-1mm}
\label{sec:splits}
\esl{Priority-aware NoCs involve complex queuing structures that cannot be modeled accurately using only the models for basic priority queuing. The complexity is primarily attributed to the partial priority contention across queues.} We identified two basic structures with partial priority dependency that constitute the building blocks of practical priority-aware NoCs. 

\begin{figure} [t]
	\centering
	\vspace{-10mm}
	\includegraphics[width=0.85\columnwidth]{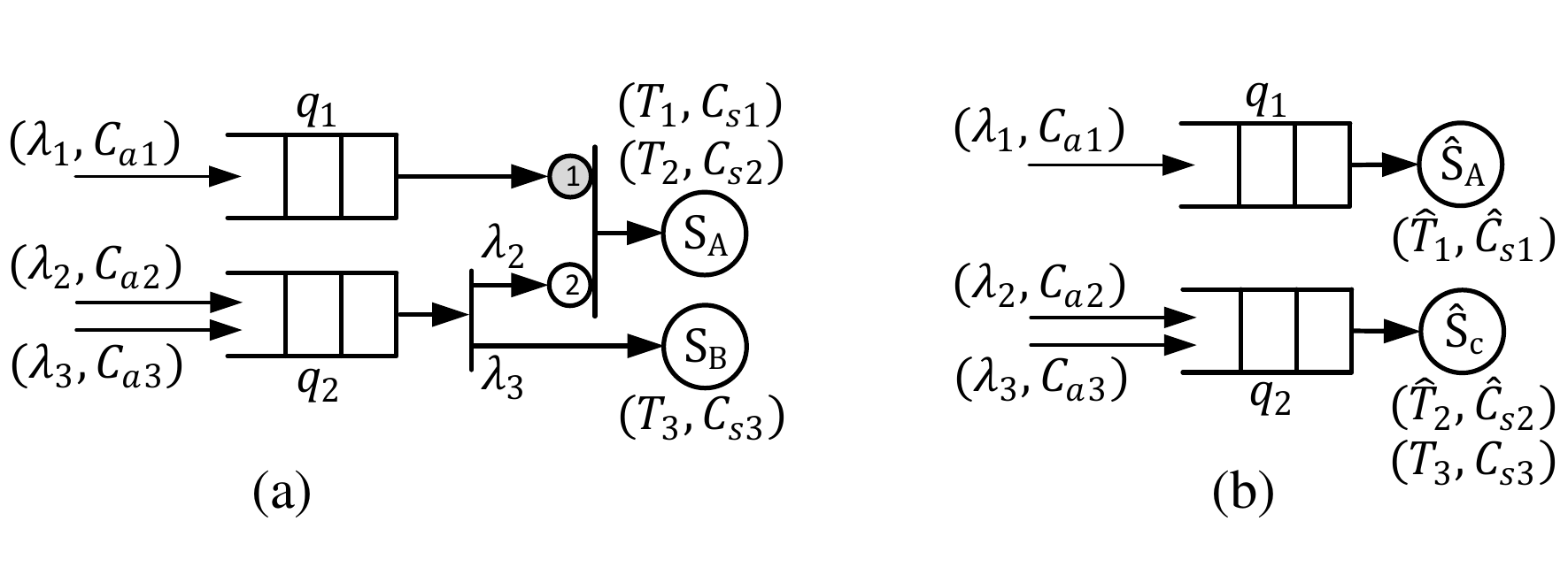}
	\vspace{-7mm}
	\centering
	\caption{Decomposition of flow contention at low priority} 
	\label{fig:split_at_low_priority_decomp}
	\vspace{-2mm}
\end{figure}

The first basic structure is shown in Figure~\ref{fig:split_at_low_priority_decomp}(a) where high priority class 1 is in contention with a portion of the traffic in $q_2$ (class 2) through server $S_A$. 
Class 2 and 3 flits have the same priority and 
share $q_2$ before entering the traffic splitter that assigns class 2 and 3 flits to 
server $S_A$ and $S_B$ respectively, following a notation similar to the one adopted in~\cite{gotmanov2011verifying}.
We denote this structure as \textit{contention at low priority}. To decompose $q_1$ and $q_2$, we need to calculate the first two moments of the modified service process of class 1 and 2.  
The decomposed structure is shown in Figure~\ref{fig:split_at_low_priority_decomp}(b). First, we set $\lambda_3$ to zero which leads to a basic priority structure. Then, we apply the decomposition method discussed in Section~\ref{basic_priority_decomp} to obtain ($\widehat{T}_1, \widehat{C}_{s_1}$) and ($\widehat{T}_2, \widehat{C}_{s_2}$). We derived mean queuing time ($W_m$) of individual classes of $q_2$ in the decomposed form as:
%
 \begin{equation} \label{eq:int_W}
     W_m = \frac{R + \sum_{k=1}^{M} \widehat{\rho}_k \widehat{T}_{k}\beta_k}{1 - \sum_{k = 1}^{M}\widehat{\rho}_{k}} + \widehat{T}_{m}(\beta_m + 1) -  T_m
 \end{equation}
%
where $R = \sum_{k = 1}^{M}\frac{1}{2}\widehat{\rho}_{k}(\widehat{T}_k - 1 + \widehat{T_k} \widehat{C}_{s_k}^2)$ and $\beta_m=\frac{1}{2} (C_{A_m}^2 + \lambda_m - 1)$.

\begin{figure} [t]
	\centering
	\vspace{-10mm}
	\includegraphics[width=1.0\columnwidth]{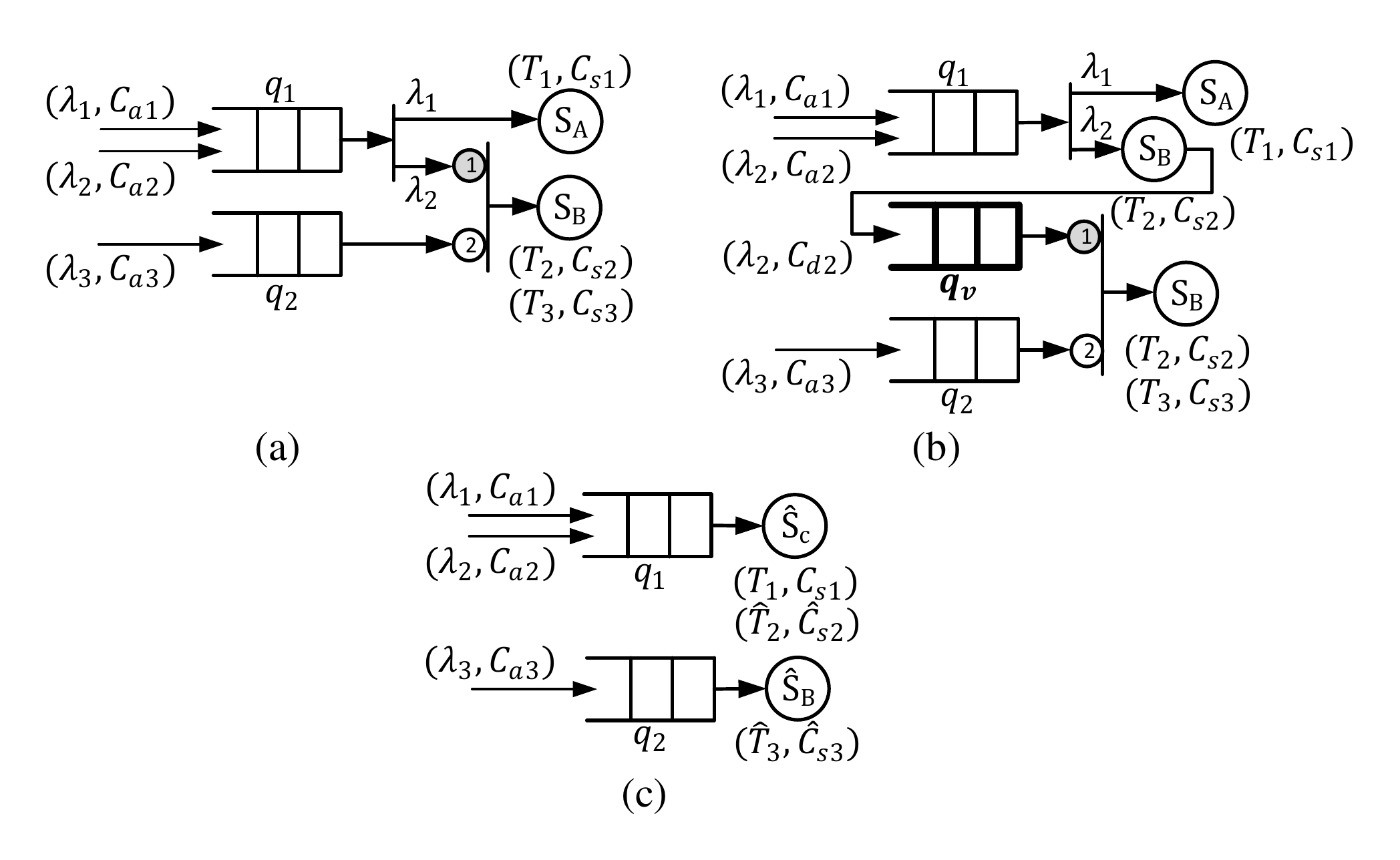}
	\vspace{-10mm}
	\centering
	\caption{Decomposition of flow contention at high priority} 
	\label{fig:split_at_high_priority_decomp}
	\vspace{-2mm}
\end{figure}

\vspace{0.5mm}
The other basic structure,~\rev{\textit{contention at high priority}, is shown in Figure~\ref{fig:split_at_high_priority_decomp}(a). In this scenario, only a fraction of the classes in $q_1$
(class 2) 
has higher priority than
class 3 since class 1 in $q_1$ is served by $S_A$.} 
\esl{Determining $\widehat{T}_3$ is challenging due to class 1 that influences the inter-departure time of class 2}.
\esl{To incorporate this effect}, we calculate the squared coefficient of variation of inter-departure time, $C_{d_2}^2$, of class 2 using the split process formulation of GGeo streams given in~\cite{kouvatsos1994entropy}. We introduce a virtual queue, $q_v$ and feed it with the flits of class 2. Therefore, $q_v$ and $q_2$ form a basic priority structure, as shown in Figure~\ref{fig:split_at_high_priority_decomp}(b). Subsequently, we apply the decomposition method described in Section~\ref{sec:basic_prior} to calculate ($\widehat{T_3}, \widehat{C}_{s_3}$) as well as ($\widehat{T_2}, \widehat{C}_{s_2}$). 
The 
decomposed structure is shown in Figure~\ref{fig:split_at_high_priority_decomp}(c).             

%


\begin{algorithm}[b]
\footnotesize
\caption{\esl{Iterative Decomposition Algorithm}} \label{algo:iter_algo}
\SetAlgoLined
\textbf{Input:} NoC topology, routing algorithm, server process, ($\lambda$) and ($p_b$) for each class as parameters \\
\textbf{Output:} Average waiting time for each queue ($W_q$) \\
$N$ = number of queues in the network\\
$S_q$ = set of classes in queue $q$\\
\For {q = \normalfont{1:$N$}}  {
\For {m = \normalfont{1:$|S_q|$}}  {
Apply decomp. for contention at high priority (if found)\\
Apply decomp. for contention at low priority (if found) \\
Compute $\widehat{T}$, $\widehat{C}_{s}^2$ using (\ref{eq:t_m}, \ref{equ:p_m(0)}) and (\ref{eq:coeff_serv}) \\
Compute queuing time ($W_{q,m}$) using (\ref{eq:int_W}) \\
}
$W_q = \frac{\sum_{i=1}^{|S_q|} \lambda_{q,m} W_{q,m}}
            {\sum_{i=1}^{|S_q|} \lambda_{q,m}}$
}
\end{algorithm}

\vspace{-4mm}
\subsection{Iterative decomposition algorithm} \label{sec:algo}
\vspace{-1mm}

\esl{Algorithm~\ref{algo:iter_algo} shows a step-by-step procedure to obtain the analytical model using our approach described in Section~\ref{sec:splits}.}
\rev{The inputs to the algorithm are NoC topology, routing algorithm and server process.
The analytical models presented for the canonical queuing system are independent of the NoC topology. Therefore, the analytical models are valid for any NoC, including irregular topologies.
}
First, we identify priority dependencies between different classes in the network. 
Next, we apply decomposition for contention at high and low priority, as shown in line 7 -- 8 of Algorithm~\ref{algo:iter_algo}.
Subsequently, we calculate the modified service process ($\widehat{T}$, $\widehat{C}_{s}^2$) using (\ref{eq:t_m}, \ref{equ:p_m(0)}) and (\ref{eq:coeff_serv}).
Then, we compute the waiting time per class following~\eqref{eq:int_W}.
Finally, we obtain the average waiting time in each queue ($W_q$), as shown in line 12.

\vspace{-4mm}
\section{\esl{Experimental Evaluation}}
\vspace{-1mm}

\begin{table*} [t]
\vspace{-10mm}
\caption{\rev{Comparisons against existing alternatives (Reference~\cite{kiasari2013analytical} 
and Reference~\cite{mandal2019analytical}). \textit{H} denotes errors over 100\%.}} 
\label{tab:error_comp}
\vspace{-3mm}
\setlength\tabcolsep{1.8pt}
\centering
\scriptsize{
\centering
\begin{adjustwidth}{-0.05cm}{}
\begin{tabular}{|l|l|l|l|l||l|l|l||l|l|l||l|l|l||l|l|l||l|l|l||l|l|l||l|l|l||l|l|l||l|l|l||l|l|l||l|l|l|} 
\hline
\multicolumn{2}{|c|}{Topology}                                                                & \multicolumn{9}{c||}{6$\times$1 Ring}                                           & \multicolumn{9}{c||}{8$\times1$ Ring}                                           & \multicolumn{9}{c||}{4$\times$4 Mesh}                                           & \multicolumn{9}{c|}{6$\times$6 Mesh}                                            \\ 
\cline{1-38}
\multicolumn{2}{|c|}{ $p_b$}                                                                  & \multicolumn{3}{c||}{0.2} & \multicolumn{3}{c||}{0.4} & \multicolumn{3}{c||}{0.6} & \multicolumn{3}{c||}{0.2} & \multicolumn{3}{c||}{0.4} & \multicolumn{3}{c||}{0.6} & \multicolumn{3}{c||}{0.2} & \multicolumn{3}{c||}{0.4} & \multicolumn{3}{c||}{0.6} & \multicolumn{3}{c||}{0.2} & \multicolumn{3}{c||}{0.4} & \multicolumn{3}{c|}{0.6}  \\ 
\cline{1-38}
\multicolumn{2}{|c|}{ $\lambda$}                                                              & 0.1 & 0.4 & 0.6          & 0.1 & 0.4 & 0.6          & 0.1 & 0.4 & 0.6          & 0.1 & 0.3 & 0.5          & 0.1 & 0.3 & 0.5          & 0.1 & 0.3 & 0.5          & 0.2 & 0.5 & 0.8          & 0.2 & 0.5 & 0.8          & 0.2 & 0.5 & 0.8          & 0.1 & 0.4 & 0.6          & 0.1 & 0.4 & 0.6          & 0.1 & 0.3 & 0.6           \\ 
\hline 
\multirow{3}{*}{\begin{tabular}[c]{@{}l@{}}\rotatebox[origin=c]{90} {Err(\%)}\\ \end{tabular}} & \textbf{Prop.} & \textbf{0.2} & \textbf{5.6} & \textbf{12}              & \textbf{0.8} & \textbf{0.6} & \textbf{12}           & \textbf{0.2} & \textbf{4.3} & \textbf{14}           & \textbf{0.5} & \textbf{3.7} & \textbf{7.3}          & \textbf{0.9} & \textbf{5.1} & \textbf{12}           & \textbf{0.5} & \textbf{3.1} & \textbf{12}              & \textbf{2.3} & \textbf{5.0} & \textbf{11}           & \textbf{2.9} & \textbf{7.5} & \textbf{13}           & \textbf{2.0} & \textbf{9.1} & \textbf{12}              & \textbf{4.7}     & \textbf{0.6}    & \textbf{11}             & \textbf{4.3} & \textbf{8.2} & \textbf{10}           & \textbf{6.1} & \textbf{7.9} & \textbf{12}            \\  
\cline{2-38}
                                                                                      & Ref\cite{kiasari2013analytical} & \textit{17}  & \birH   & \birH              & \textit{30}  & \birH   & \birH            & \textit{54}  & \birH   & \birH            & \textit{66}  & \birH   & \birH            & \birH   & \birH   & \birH            & \birH   & \birH   & \birH              & \textit{30}  & \birH   & \birH            & \textit{10}  & \birH   & \birH            & \textit{12}  & \birH   & \birH              & \textit{28}     & \birH    & \birH             & \textit{54}  & \birH   & \birH            & \textit{78}  & \birH   & \birH             \\
                                                                                      \cline{2-38}
                                                                                      & \cellcolor{LightCyan}  Ref\cite{mandal2019analytical}   & 8.5 & 12  & 18              & 20  & 30  & 55           & 36  & 47  & 79           & 7.5 & 8.8 & 11           & 18  & 24  & 39           & 33  & 42  & 85              & 10  & 21  & 40           & 21  & 38  & 82           & 37  & 56  & 88              &7.2     &13     & 45             & 14  & 34  & 64           & 28  & 48  & 76            \\
\hline
\end{tabular}
\end{adjustwidth}}
\vspace{-1mm}
\end{table*}

The proposed technique is implemented in C++ to facilitate integration with system-level simulators. 
\camready{Analysis takes 2.7~ms for a 6$\times$6 NoC and the worst-case complexity is $O(n^3)$, where $n$ is the number of nodes.}
\camready{In all experiments, 200K cycles of warm-up period is considered.}
The accuracy of the models is evaluated against an industrial cycle-accurate simulator~\cite{ogras2012energy}
under both real applications and synthetic traffic that models
uniformly distributed core to last-level cache traffic with 100\% hit rate. 

%



\vspace{-4mm}
\subsection{\esl{Evaluation on Architectures with Ring NoCs}}
\vspace{-1mm}

This section analyzes the accuracy of the proposed analytical models using uniform traffic 
on a priority-based 6$\times$1 and 8$\times$1 ring NoCs, 
similar to those used in high-end client CPUs with integrated GPU and memory controller. 
\rev{Table~\ref{tab:error_comp} shows that} the average errors between our technique and simulation are 
6\%, 4\% and 6\% for burst probability of 0.2, 0.4 and 0.6, respectively.
These errors hardly reach 14\% even at the highest injection, which is hard to model. 
Table~\ref{tab:error_comp} also shows that priority-based analytical models \textit{which do not} consider burstiness~\cite{mandal2019analytical} significantly underestimate the latency by 33\% on average (highlighted with the shaded row). 
\rev{In contrast, the work without the proposed decomposition technique~\cite{kiasari2013analytical} leads to over 100\% overestimation even at low traffic loads (highlighted with text in italics).}
\esl{In this case, GGeo models can not handle partial contention, since it assumes all packets in the high-priority queue have higher priority than each packet in the low priority queue.}
These results demonstrate that the proposed priority-aware NoC performance models have significantly higher accuracy than the existing alternatives.

%
%
%
\begin{figure}[b]
	\centering
	\vspace{-1mm}
	\includegraphics[width=0.85\columnwidth]{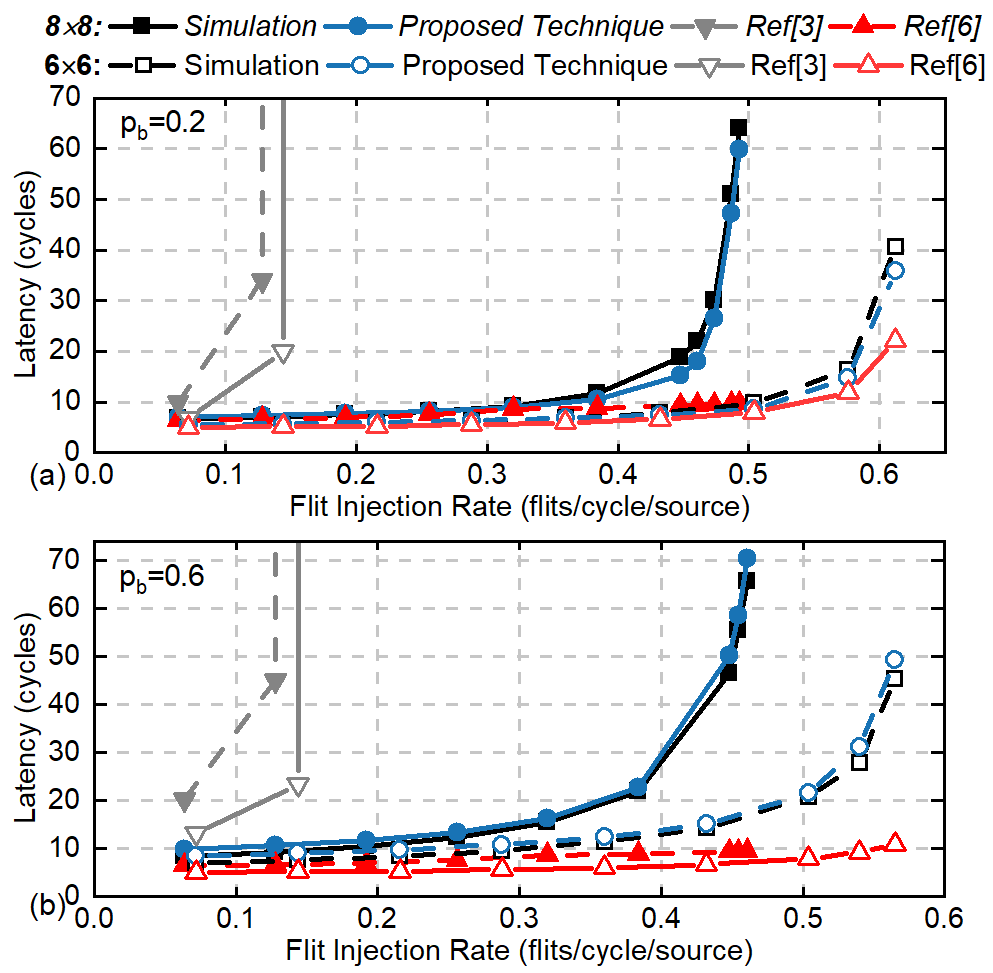}
	\centering
	\vspace{-4mm}
	\caption{\rev{Comparison of a proposed analytical model with cycle-accurate simulation for 8$\times$8 and 6$\times$6 mesh for (a) $p_b = 0.2$ and (b) $p_b = 0.6$.}}
	\label{fig:mesh}
	\vspace{-9mm}
\end{figure}
\vspace{-4mm}
\subsection{\esl{Evaluation on Architectures with Mesh NoCs}}
\vspace{-1mm}
Table~\ref{tab:error_comp} compares the analytical model and simulation results for a priority-based 4$\times$4 and 6$\times$6 mesh NoC, similar to those used in high-end servers~\cite{jeffers2016intel}. 
Our technique incurs on average 6\%, 7\% and 10\% error for burst probability of 0.2, 0.4 and 0.6, respectively.
Priority-based analytical models which neglect burstiness~\cite{mandal2019analytical} underestimate the latency by 60\% on average similar to the results on the ring architectures.
Likewise, GGeo models without the proposed decomposition technique lead to overestimation.
\rev{We also provide detailed comparison of proposed analytical models on 6$\times$6 and 8$\times$8 NoC for burst probability of 0.2 and 0.6 in Figure~\ref{fig:mesh}(a) and Figure~\ref{fig:mesh}(b), respectively. The proposed models significantly outperform the other alternatives and 
lead to less than 10\% error on average.} 


\vspace{-5mm}
\subsection{Evaluation with Real Applications}
\vspace{-1mm}
\rev{This section validates the proposed analytical models 
using SYSmark\textsuperscript{\textregistered} 2014 SE~\cite{SYSmark2014},
and applications from SPEC CPU\textsuperscript{\textregistered} 2006~\cite{henning2006spec} and SPEC CPU\textsuperscript{\textregistered} 2017~\cite{bucek2018spec} benchmark suites.} 
These applications are chosen since they show different levels of burstiness.
First, we run these applications on gem5~\cite{Binkert2011Gem5}
and collect traces with timestamps for each packet injection.
\esl{Then, we use the traces to compute the injection rate ($\lambda$) and $p_b$.}

\noindent\esl{\textbf{Computing $p_b$}: 
For each source, we feed traffic arrivals with timestamps over a 200K clock cycle window into a virtual queue with the same service rate as the NoC to determine the queue occupancy. 
At the end of the window, we compute the average occupancy.
Then, we employ the model described in~\cite{kouvatsos1994entropy} 
to find the occupancy and then $p_b$ of each class}.

The proposed analytical models are used to estimate the \camready{latency} using the injection rate and burst parameters, as well as the NoC architecture and routing algorithm.
The applications show burstiness in the range of 0.2 -- 0.5.
\rev{As shown in Table~\ref{tab:real_app}, the proposed technique has on average 2\% and 4\% error compared to cycle-accurate simulations for 6$\times$6 mesh and 8$\times$8 mesh, respectively}.
\rev{In contrast, the analytical models presented in~\cite{kiasari2013analytical} and~\cite{mandal2019analytical} incur significant modeling error.}


\begin{table}[h]
\vspace{-3mm}
\caption{\rev{Modeling Error (\%) with Real Applications}} \label{tab:real_app}
\vspace{-3mm}
\setlength\tabcolsep{1.8pt}
\centering
\scriptsize{
\begin{tabular}{|c|c|c|c|c|c|c|c|c|c|}
\hline
                                                                      &                                & {  \begin{tabular}[c]{@{}l@{}}xalan-\\ cbmk\end{tabular}} & {  mcf}   & {  gcc}   & {  bwaves} & {  \begin{tabular}[c]{@{}l@{}}Gems\\ FDTD\end{tabular}} & {  \begin{tabular}[c]{@{}l@{}}omnet-\\ pp\end{tabular}} & {  \begin{tabular}[c]{@{}l@{}}perl-\\ bench\end{tabular}} & {  \begin{tabular}[c]{@{}l@{}}SYSmark\\ 14se\end{tabular}} \\ \hline
                                                                      & {  Prop}       & {  2.17}                                                  & {  4.97}  & {  0.92}  & {  0.15}   & {  0.38}                                                & {  5.10}                                                & {  3.63}                                                  & {  0.73}                                                   \\ \cline{2-10} 
                                                                      & {  Ref~\cite{kiasari2013analytical}} & {  14.62}                                                 & {  11.99} & {  7.69}  & {  12.29}  & {  5.18}                                                & {  13.64}                                               & {  11.46}                                                 & {  7.25}                                                   \\ \cline{2-10} 
\multirow{-3}{*}{\begin{tabular}[c]{@{}l@{}}6$\times$6 \\ Mesh\end{tabular}} & {  Ref~\cite{mandal2019analytical}} & {  17.36}                                                 & {  23.29} & {  7.71}  & {  22.02}  & {  6.99}                                                & {  14.11}                                               & {  12.95}                                                 & {  11.13}                                                  \\ \hline
                                                                      & {  Prop}       & {  3.59}                                                  & {  4.08}  & {  3.81}  & {  4.87}   & {  0.44}                                                & {  7.48}                                                & {  3.67}                                                  & {  1.10}                                                   \\ \cline{2-10} 
                                                                      & {  Ref~\cite{kiasari2013analytical}} & {  10.33}                                                 & {  12.73} & {  12.07} & {  22.90}  & {  19.17}                                               & {  9.93}                                                & {  5.99}                                                  & {  19.04}                                                  \\ \cline{2-10} 
\multirow{-3}{*}{\begin{tabular}[c]{@{}l@{}}8$\times$8\\ Mesh\end{tabular}}  & {  Ref~\cite{mandal2019analytical}} & {  12.15}                                                 & {  29.99} & {  10.00} & {  19.65}  & {  5.44}                                                & {  10.78}                                               & {  14.74}                                                 & {  7.94}                                                   \\ \hline
\end{tabular}}
\vspace{-4mm}
\end{table}

\vspace{-2mm}
\section{Conclusion}

We presented analytical models for priority-aware NoCs under bursty traffic.
We model bursty traffic as generalized geometric distribution and applied the maximum entropy method to construct analytical models.
Experimental evaluations 
show that the proposed technique \esl{has less 10\% modeling error with respect to cycle-accurate NoC simulation for real applications.}
\vspace{-4mm}
   



 \section*{Appendix A}

\begin{figure}[h]
 	\centering
 	\includegraphics[width=1\columnwidth]{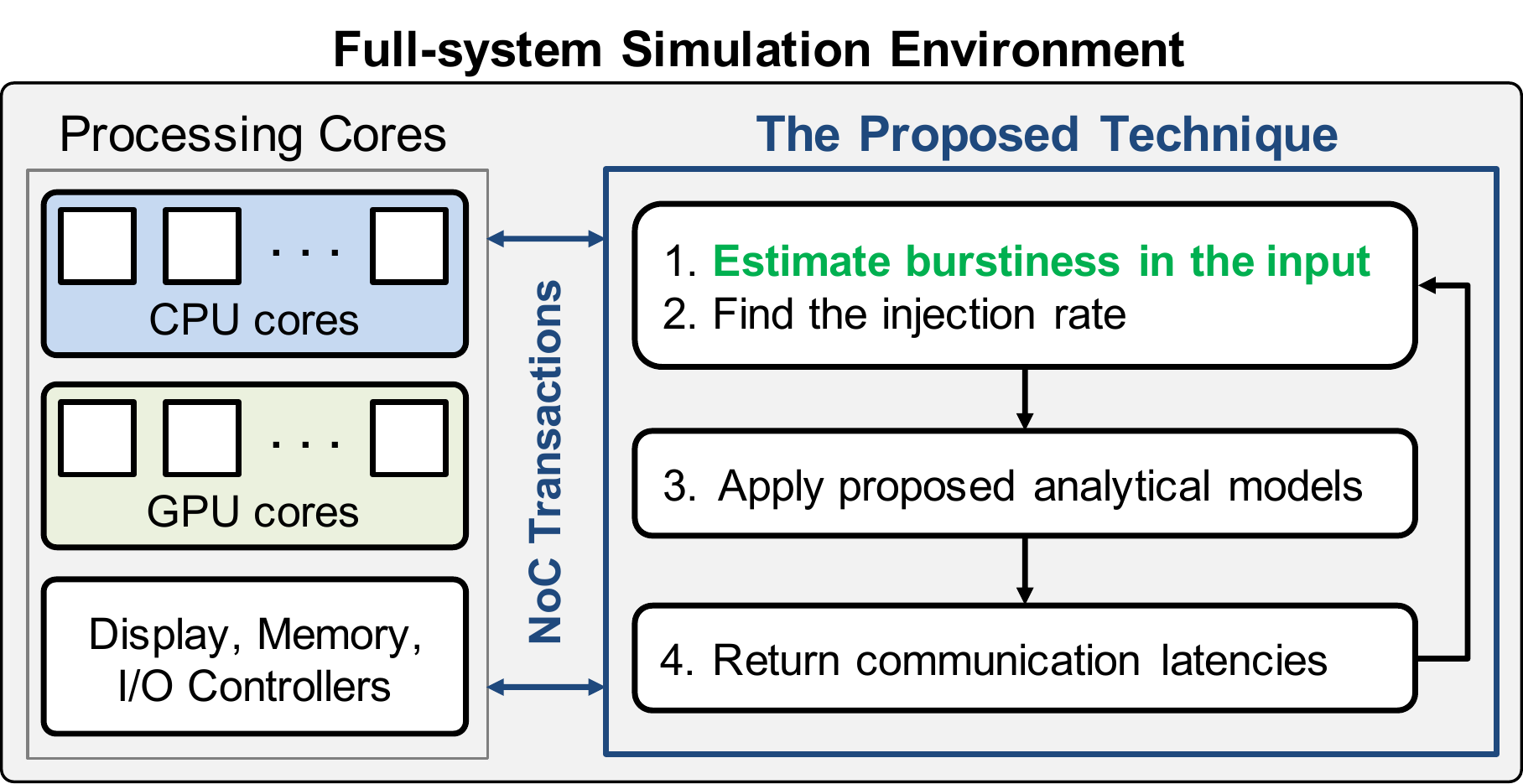}
 	\centering
 	\caption{An overview of the proposed approach} 
 	\label{fig:overview}
 \end{figure}

 \noindent\textbf{Usage of the proposed analytical models:}
 In this work, we aim to replace the cycle accurate NoC simulators with analytical performance models.
 The full-system simulation environment keeps track of the traffic injected from processing cores (e.g. CPU, GPU, caches, memory etc.) to the NoC, as shown in Figure~\ref{fig:overview}.
 The proposed technique obtains the traffic information of processing cores over a time window, which is in the order of 100-200K cycles in our experiments.  
 The duration can be decreased if the workload characteristics change considerably within a window or increased if the workload is steady. 
 Our simulator estimates the burstiness of the input and calculates the injection rate of each traffic class using this information (first two steps in Figure~\ref{fig:overview}).
 Then, it applies the proposed analytical models to obtain end-to-end latency of each traffic class.
 Whenever a processing core issues a new transaction, the communication latency is computed using these models instead of cycle-accurate simulations. These steps are repeated for each time window.

 \section*{Appendix B}

 \noindent\textbf{Generalization of the proposed analytical models:}
 We incorporate Y-X routing algorithm in the experimental evaluations presented following the actual reference commercial hardware design~\cite{jeffers2016intel}. However, we note that the proposed approach is independent of the routing algorithm. In fact, the routing algorithm is one of the inputs to the proposed Iterative Decomposition Algorithm (Algorithm 1).

 The analytical models are valid for any type of NoC including irregular topologies. The analytical models presented for the canonical queuing system are independent of the NoC topology. The canonical model constitutes the end-to-end latency model for a given NoC topology. In fact, NoC topology is an input to the algorithm which computes the end-to-end latency (Algorithm 1). Since this work targets general purpose NoCs used in manycore processors, we evaluate our proposed model only with Mesh and Ring NoC used in Intel Xeon server~\cite{jeffers2016intel}, Xeon Phi~\cite{sodani2016knights}, and quad-core i7 (with integrated graphics)~\cite{charles2009evaluation} processors.

 \begin{table}[t]
 \caption{Probability of burstiness ($p_{b}$) for different applications.} 
 \label{tab:app_burst}
 \setlength\tabcolsep{2pt}
 \centering
 \begin{tabular}{@{}c|cccccccc@{}}
 \toprule
 Apps & \begin{tabular}[c]{@{}c@{}}xalan-\\ cbmk\end{tabular} & mcf  & gcc & bwaves  & \begin{tabular}[c]{@{}c@{}}Gems-\\FDTD\end{tabular} & \begin{tabular}[c]{@{}c@{}}omnet-\\pp\end{tabular} & \begin{tabular}[c]{@{}c@{}}perl-\\bench\end{tabular} & \begin{tabular}[c]{@{}c@{}}SYSmark-\\ 14se\end{tabular} \\ \midrule
 $p_{b}$                                              & 0.37                                                  & 0.43 & 0.26   & 0.53 & 0.26                                                & 0.18    & 0.26  & 0.27                                                  \\ \bottomrule
 \end{tabular}
 \end{table}

 \begin{table}[b]
 \caption{Modeling Error (\%) with Real Applications} \label{tab:real_app_6x1}
 \setlength\tabcolsep{1.8pt}
 \centering
 \scriptsize{
 \begin{tabular}{|c|c|c|c|c|c|c|c|c|c|}
 \hline
                                                                       &                                & {  \begin{tabular}[c]{@{}l@{}}xalan-\\ cbmk\end{tabular}} & {  mcf}   & {  gcc}   & {  bwaves} & {  \begin{tabular}[c]{@{}l@{}}Gems\\ FDTD\end{tabular}} & {  \begin{tabular}[c]{@{}l@{}}omnet-\\ pp\end{tabular}} & {  \begin{tabular}[c]{@{}l@{}}perl-\\ bench\end{tabular}} & {  \begin{tabular}[c]{@{}l@{}}SYSmark\\ 14se\end{tabular}} \\ \hline
                                                                       & {  Prop}       & {  5.38}                                                  & {  5.48}  & {  4.57}  & {  6.94}   & {  0.31}                                                & {  6.58}                                                & {  9.91}                                                  & {  1.20}                                                   \\ \cline{2-10} 
                                                                       & {  Ref~\cite{kiasari2013analytical}} & {  24.06}                                                 & {  13.20} & {  11.80} & {  19.89}  & {  9.79}                                                & {  10.94}                                               & {  13.53}                                                 & {  7.74}                                                   \\ \cline{2-10}
 \multirow{-3}{*}{\begin{tabular}[c]{@{}l@{}}6$\times$1 \\ Ring\end{tabular}} & {  Ref~\cite{mandal2019analytical}} & {  27.14}                                                 & {  17.29} & {  19.35} & {  29.40}  & {  12.54}                                               & {  10.74}                                               & {  20.26}                                                 & {  10.24}                                                  \\ \hline
        
 \end{tabular}
 }
 \end{table}

 \section*{Appendix C}

 \noindent\textbf{Results with real application executed on 6$\times$1 ring:}
 Table~\ref{tab:app_burst} shows the probability of burstiness of different applications used in our experiemntal evaluations.
 The levels of burstiness exhibited by these applications are between 0.18 and 0.53 
 re-emphasizing that the chosen levels of burstiness for evaluation with synthetic traffic in Section V-C are representative of real applications.

 Table~\ref{tab:real_app_6x1} shows the modeling error with respect to simulation for 6$\times$1 ring for the proposed approach, the approach presented in~\cite{kiasari2013analytical} and~\cite{mandal2019analytical}.
 The error with proposed analytical model is always less than 10\%.
 However, the technique presented in~\cite{kiasari2013analytical} does not take care multiple traffic classes in single queue resulting up to 24\% error with respect to cycle-accurate simulation.
 Moreover, the analytical model constructed in~\cite{mandal2019analytical} does not incorporate burstiness of the input traffic which results in upto 28\% modeling error.




\bibliographystyle{unsrt}
\small{
\bibliography{main}
}



\end{document}